\documentclass[bimj,fleqn]{w-art}
\usepackage{times}
\usepackage{w-thm}
\usepackage[authoryear]{natbib}
\theoremstyle{plain}

\theoremstyle{definition}

\usepackage[]{graphicx}
\chardef\bslash=`\\ % p. 424, TeXbook

\usepackage{booktabs}  % for nicer tables
\usepackage{verbatim}  % for "\verbatiminput{}"
\usepackage{hyperref}
\providecommand{\normaldistn}{\mathrm{Normal}}
\providecommand{\expect}{\mathrm{E}}

\begin{document}
%\DOIsuffix{bimj.200100000}
\DOIsuffix{bimj.DOIsuffix}
\Volume{XX}
\Issue{YY}
\Year{2020}
\pagespan{1}{}
  %%% semicolon and fullpoint added here for keyword style

\title{Bounds for the weight of external data in shrinkage estimation}
%% Information for the first author.
\author[C.~R\"{o}ver]{Christian R\"{o}ver\footnote{Corresponding author: {\sf{e-mail: christian.roever@med.uni-goettingen.de}}}\inst{,1}} 
\address[\inst{1}]{Department of Medical Statistics, University Medical Center G\"{o}ttingen, Humboldtallee~32, 37073~G\"{o}ttingen, Germany}
%%%%    Information for the second author
\author[T.~Friede]{Tim Friede\inst{1}}
%%%%    \dedicatory{This is a dedicatory.}
\Receiveddate{xxx} \Reviseddate{yyy} \Accepteddate{zzz} 

% ORCID iDs:
%  Christian Röver http://orcid.org/0000-0002-6911-698X
%  Tim Friede      http://orcid.org/0000-0001-5347-7441

\begin{abstract}
  Shrinkage estimation in a meta-analysis framework may be used to
  facilitate dynamical borrowing of information.
  This framework might be used to analyze a new study in the light of
  previous data, which might differ in their design (e.g., a
  randomized controlled trial (RCT) and a clinical registry).
  We show how the common study weights arise in effect and shrinkage
  estimation, and how these may be generalized to the case of Bayesian
  meta-analysis. Next we develop simple ways to compute bounds on the
  weights, so that the contribution of the external evidence may be
  assessed \emph{a~priori}. These considerations are illustrated and
  discussed using numerical examples, including applications in the
  treatment of Creutzfeldt-Jakob disease and in fetal monitoring to
  prevent the occurrence of metabolic acidosis.  The target study's
  contribution to the resulting estimate is shown to be bounded below.
  Therefore, concerns of evidence being easily overwhelmed by external
  data are largely unwarranted.
\end{abstract}

\maketitle

%%%%%%%%%%%%%%%%%%%%%%%%%%%%%%%%%%%%%%%%%%%%%%%%%%%%%%%%%%%%%%%%%%%%%%%%%%%%%%%%
  \section{Introduction}

    In some situations it is useful to support an estimate using
    additional external evidence, for example, when a small study in
    the context of a rare disease may be supplemented with data from a
    clinical registry or electronic health records, or when the result
    from a meta-analysis may be backed by an analysis in a similar
    field, e.g., a related but somewhat different population.  The
    involved data contributions then take on different roles, namely,
    that of a \emph{source} (the external data) and a \emph{target}
    (the data of primary interest).  \emph{Dynamic borrowing} refers
    to the class of approaches where the apparent, empirical
    similarity or compatibility of the source and the target is taken
    into account when judging to what degree the two should be lumped
    together \citep{RoeverFriede2020}.  Such approaches may be
    implemented, e.g., via hierarchical models or informative priors;
    both are actually equivalent to some degree in the context of the
    \emph{normal-normal hierarchical model (NNHM)}
    \citep{SchmidliEtAl2014}. Similarly, closely related (or partly
    equivalent) approaches are given by the \emph{bias allowance}
    framework \citep{WeltonEtAl} or the \emph{power prior} framework
    \citep{IbrahimChen2000}.  A recent example of such an approach is
    given by the \textsc{early pro-tect} trial in paediatric Alport
    disease, where data from a randomized controlled trial (RCT) were
    supported by source data from an open-label arm and a clinical
    registry \citep{GrossEtAl2020}.

    In the context of dynamic borrowing within the NNHM framework, the
    flow of information is quite commonly illustrated by quoting
    \emph{weights} of data sources as these are combined to a joint
    estimate. As the eventual estimate may be expressed as a weighted
    average of the input data, the corresponding weights are a useful
    means of quantifying the studies' contributions to or influence on
    the eventual result \citep{HedgesOlkin,HartungKnappSinha}.
    Analogous weights arise for shrinkage estimates
    \citep{RaudenbushBryk1985,Robinson1991,Viechtbauer2010}, and, as
    we will show below, also in the Bayesian paradigm with prior
    distributions on effect and heterogeneity parameters.

    When combining originally separate data sets in a meta-analysis or
    using shrinkage estimation, there sometimes is concern that
    evidence from the target data may be overwhelmed by a much larger
    set of source data, e.g., when combining a small RCT with a large
    clinical registry or routine data (e.g. electronic health records)
    \citep{WeberHemmingsKoch2018}.  In such cases it is instructive to
    explicate the notion of study contributions by considering their
    weights.  Again, we can see the dynamic nature of the approach in
    the changing weight of external data with varying data
    compatibility or discrepancy. It turns out that within the
    Bayesian framework we can determine the \emph{minimum weight} of
    the target study (the RCT in the above example) a~priori for a
    given analysis, and with that we are able to provide more insights
    into the general behaviour of the meta-analysis procedure. The
    derived formulas show shrinkage estimation to behave reasonably
    and also predictably.

    In the following, we will review the NNHM, and show how ``study
    weights'' arise in effect and shrinkage estimation and how the
    concept may be extended to the Bayesian framework. Then we take a
    closer look at the weights' properties and show how these may be
    bounded across possible prior settings and/or data
    realisations. The arguments are illustrated by a numerical study,
    and the ideas are employed in two example applications involving
    the joint analysis of a ``small'' target and a ``large'' source
    study, as well as two equally-sized studies.  Due to the
    few-studies setup
    \citep{FriedeRoeverWandelNeuenschwander2017b,RoeverFriede2020}, we
    will be focusing on Bayesian methods and only in between point out
    some connections to common analogous frequentist results.  We
    close with a discussion of the findings and their practical
    implications.

%%%%%%%%%%%%%%%%%%%%%%%%%%%%%%%%%%%%%%%%%%%%%%%%%%%%%%%%%%%%%%%%%%%%%%%%%%%%%%%%
  \section{The normal-normal hierarchical model (NNHM)}\label{sec:nnhm}
    The NNHM models a set of $k$~estimates~$y_i$ and their standard
    errors~$\sigma_i$ as
    \begin{equation}\label{eqn:NNHM1}
      y_i|\theta_i,\sigma_i \; \sim \; \normaldistn(\theta_i, \sigma_i^2)\mbox{,}
    \end{equation}
    where $\theta_i$~are the \emph{study-specific effects}. The
    $\theta_i$~are not necessarily identical for all studies, but they
    are also associated with a certain amount of variation, expressed
    as
    \begin{equation}\label{eqn:NNHM2}
      \theta_i|\mu,\tau \; \sim \; \normaldistn(\mu,\tau^2)\mbox{.}
    \end{equation}
    The mean parameter~$\mu$ is the \emph{overall mean effect}, while
    $\tau$ denotes the \emph{between-study variability
      (heterogeneity)}.  As noted elsewhere
    \citep{HedgesOlkin,HartungKnappSinha,Roever2020}, marginalizing
    over the parameters $\theta_i$, the model may be written as
    \begin{eqnarray}\label{eqn:NNHM3}
      y_i|\mu,\tau,\sigma_i & \sim & \normaldistn(\mu,\,\sigma_i^2+\tau^2)\mbox{.} \label{eqn:NNHM}
    \end{eqnarray}
    The NNHM is a random-effects (RE) model, which in the special case
    of~$\tau=0$ reduces to a fixed-effect (FE) (or common-effect)
    model.  
    It provides a good approximation for many types of effect measures
    where measurement uncertainty and between-study variability may be
    assumed to be (approximately) normally distributed
    \citep{JacksonWhite2018}.
    Data analysis may then aim at estimating the overall
    effect~$\mu$ or study-specific effects~$\theta_i$ (``shrinkage
    estimation''); in the present investigation, we will mostly be
    concerned with the latter.

    In the following, we will denote vectors of effect estimates
    $(y_1,\ldots,y_k)$ and their standard errors
    $(\sigma_1,\ldots,\sigma_k)$ by~$\vec{y}$ and~$\vec{\sigma}$,
    respectively. Furthermore, we will be mostly concerned with the
    special case of only two studies ($k=2$) and a non-informative
    (improper) uniform prior for the overall effect ($p(\mu) \propto
    1$).

%%%%%%%%%%%%%%%%%%%%%%%%%%%%%%%%%%%%%%%%%%%%%%%%%%%%%%%%%%%%%%%%%%%%%%%%%%%%%%%%
  \section{Study weights}\label{sec:weights}
    \subsection{Conditional weights}
      Assuming an (improper) uniform prior for the overall
      effect~$\mu$, the conditional posterior distribution of~$\mu$
      (given $\tau$) is normal with mean
      \begin{equation}\label{eqn:overallMean01}
        \tilde{\mu}(\tau)
        \;=\; \expect[\mu|\tau,\vec{y},\vec{\sigma}]
        \;=\; \sum_{i=1}^k w_i(\tau)\, y_i \mbox{,}
      \end{equation}
      where the \emph{inverse-variance (IV) weights} $w_j(\tau)$
      are given by
      \begin{equation}
        w_j(\tau)
        \;=\; \frac{\frac{1}{\sigma_j^2+\tau^2}}{\sum_{i=1}^k \frac{1}{\sigma_i^2+\tau^2}}
      \end{equation}
      as in the frequentist framework
      \citep{HedgesOlkin,HartungKnappSinha,FriedeRoeverWandelNeuenschwander2017a}.
      A similar formula also applies for a normal prior
      \citep{Roever2020}.
%     Roever (2017; Sec. 2.4)
      These two (conditionally conjugate) priors are computationally
      simple, readily motivated, and because of that reason, probably
      also the most commonly used priors for~$\mu$ for this model.
      Informative priors might for example be motivated by general
      plausibility considerations or empirical data
      \citep{GunhanRoeverFriede2020}, or determined through expert
      elicitation \citep{HampsonEtAl2014,BestDallowMontague2020}.

      The conditional posterior of the study-specific
      effect~$\theta_j$ (the \emph{shrinkage estimate}) is also normal
      with mean~$\tilde{\theta}_j(\tau)$ depending on~$y_i$
      and~$\tilde{\mu}(\tau)$, namely
      \begin{equation}\label{eqn:shrinkMean01}
        \tilde{\theta}_j(\tau)
        \;=\; \expect[\theta_j|\tau,\vec{y},\vec{\sigma}]
        \;=\; b_j(\tau)\, y_j + \bigl(1-b_j(\tau)\bigr)\, \tilde{\mu}(\tau) 
      \end{equation}
      where the corresponding
      weight \citep{Roever2020,WandelNeuenschwanderRoeverFriede2017} is
      \begin{equation}
        b_j(\tau)
        \;=\; \frac{\frac{1}{\sigma_j^2}}{\frac{1}{\sigma_j^2}+\frac{1}{\tau^2}}\mbox{.}
      \end{equation}
      The formulation in (\ref{eqn:shrinkMean01}) shows to which
      degree the estimate is \emph{shrunk} towards the common overall
      mean~$\tilde{\mu}(\tau)$ (depending on the amount of
      heterogeneity).  Equation~(\ref{eqn:shrinkMean01}) may be
      re-written as
      \begin{eqnarray}
        \tilde{\theta}_j(\tau)
        &=&  \Bigl[b_j(\tau) + \bigl(1-b_j(\tau)\bigr)w_j(\tau)\Bigr] \, y_j 
             \,+\,  \sum_{i \neq j}\Bigl[ \bigl(1-b_j(\tau)\bigr) \, w_i(\tau)\Bigr] \, y_i\\
        &=& c_{jj}(\tau)\,y_j + \sum_{i\neq j}c_{ij}(\tau)\,y_i\\
        &=& \sum_{i=1}^k c_{ij}(\tau)\,y_i  \label{eqn:shrinkMean02}
      \end{eqnarray}
      so that the actual \emph{shrinkage weights}~$c_{ij}(\tau)$ (the
      weight of the $i$th study in the estimate of the $j$th study)
      become more explicit.  In the special case of only two studies
      ($k=2$), the coefficients~$c_{ij}(\tau)$ simplify to
      \begin{equation}
        c_{11}(\tau)\;=\;\frac{\sigma_2^2+2\tau^2}{\sigma_1^2+\sigma_2^2+2\tau^2}
        \mbox{,}\qquad
        c_{12}(\tau)\;=\;\frac{\sigma_1^2}{\sigma_1^2+\sigma_2^2+2\tau^2}
        \mbox{,} \label{eqn:k2ShrinkWeights}
      \end{equation}
      and analogously for~$c_{22}$ and~$c_{21}$.

      The conditional mean~$\tilde{\mu}(\tau)$ commonly also arises in
      frequentist approaches as an overall effect estimator, where
      usually a heterogeneity estimate~$\hat{\tau}$ is plugged in
      for~$\tau$ \citep{HedgesOlkin,HartungKnappSinha}.  Similarly,
      plug-in estimates of~$\tilde{\theta}_j(\tau)$ are widely used
      and commonly known as ``best linear unbiased prediction (BLUP)''
      \citep{RaudenbushBryk1985,Robinson1991,Viechtbauer2010}.  The
      weights~($w_j(\tau)$ or~$c_{ij}(\tau)$) are then often quoted
      along with the results in order to illustrate the individual
      studies' contributions to the overall result.  Note that while
      weights may be appealing, they still constitute an ultimately
      somewhat heuristic notion of the concept of \emph{a study's
        contribution}, as these only relate to the posterior
      expectation.  A more complete picture might actually be obtained
      by considering the corresponding meta-analytic-predictive (MAP)
      prior \citep{SchmidliEtAl2014}, which comprehensively describes
      the information conveyed by the source study.

    \subsection{Marginal weights}
      In a Bayesian multiparameter model, the \emph{conditional}
      expectations (of effects~$\mu$ or~$\theta_j$) as derived above
      are commonly of limited interest; what is usually more
      interesting are the \emph{marginal} posterior expectations, as
      these refer to the posterior distribution integrated over other
      parameters such as the heterogeneity~$\tau$ in the considered
      model.  Marginal posterior expectations here result from the
      conditional expectations as expected values with respect to the
      heterogeneity's marginal posterior
      distribution~$p(\tau|\vec{y},\vec{\sigma})$, i.e.,
      \begin{equation}
        \expect[\mu|\vec{y},\vec{\sigma}]
        =  \expect_{p(\tau|\vec{y}, \vec{\sigma})} \Bigl[ \expect[\mu|\tau, \vec{y}, \vec{\sigma}] \Bigr]
        \; \mbox{ and } \;
        \expect[\theta_j|\vec{y},\vec{\sigma}]
        =  \expect_{p(\tau|\vec{y}, \vec{\sigma})} \Bigl[ \expect[\theta_j|\tau, \vec{y}, \vec{\sigma}] \Bigr]
           \mbox{.}
      \end{equation}
      In both cases, the conditional expectations result in convex
      combinations of the form $\sum_i\alpha_i(\tau)\,y_i$ (see
      Equations~(\ref{eqn:overallMean01})
      and~(\ref{eqn:shrinkMean02})).  For convex (or, more generally,
      linear) combinations, we may re-write the expectations as
      \begin{equation}
        \expect_{p(\tau|\vec{y}, \vec{\sigma})} \Bigl[\sum_i\alpha_i(\tau)\,y_i\Bigr]
        \;=\; \sum_i \expect_{p(\tau|\vec{y}, \vec{\sigma})} [\alpha_i(\tau)] \,y_i
        \mbox{,}
      \end{equation}
      so that it becomes apparent that the marginal expectation may
      again be expressed as a weighted average of the effects~$y_i$,
      where the study weights now arise as the \emph{posterior
        expected weights}.  These constitute straightforward
      generalizations of the common \emph{conditional} weights to the
      Bayesian context.  The weights can be obtained from
      one-dimensional integrals (expectations) involving the
      heterogeneity's marginal posterior distribution and may easily
      be computed numerically; they are returned by default by the
      \texttt{bayesmeta} \textsf{R}~package
      \citep{bayesmeta,Roever2020}.

    \subsection{Properties}
      For $\tau=0$ the NNHM reduces to the FE model, in which all
      study effects~$\theta_i$ coincide with the overall mean~$\mu$.
      As $\tau$ is varied between the two extremes of $\tau=0$ and
      $\tau\rightarrow\infty$, several effects may be observed for the
      conditional weights:
      \begin{itemize}
        \item The IV-weights~$w_j(\tau)$ move (not necessarily
          monotonically) from ``fixed-effect'' weights
          $w_j(0)=\frac{\frac{1}{\sigma_j^2}}{\sum_i\frac{1}{\sigma_i^2}}$
          that depend on the study's precision towards ``average''
          weights $w_i(\infty)=\frac{1}{k}$ where all studies have the
          same weight.
        \item The weights~$b_j(\tau)$ increase monotonically from~$0$
          towards~$1$.
        \item The shrinkage weights~$c_{jj}(\tau)$ (the contribution
          of the $j$th~study to its own shrinkage estimate) increase
          monotonically from the FE weight towards~$1$.
      \end{itemize}
      For the conditional expectations, this implies:
      \begin{itemize}
        \item The conditional effect estimate~$\tilde{\mu}(\tau)$
          moves from the FE estimate towards an unweighted average.
        \item The conditional shrinkage
          estimates~$\tilde{\theta}_j(\tau)$ move from the FE estimate
          towards the ``un-pooled'' original estimates~$y_j$.
      \end{itemize}
      These effects are also summarized in
      Table~\ref{tab:VaryingTau}. Posterior expectations of the
      weights depend on the heterogeneity's posterior
      distribution~$p(\tau|\vec{y}, \vec{\sigma})$.  For a uniform
      effect prior, a given heterogeneity prior~$p(\tau)$ and standard
      errors~$\sigma_i$, the posterior density is given by
      \begin{equation}
        p(\tau|\vec{y},\vec{\sigma})
        \; \propto \; p(\tau) \, f_{\vec{\sigma}}(\tau) \, g_{\vec{y}}(\tau) \label{eqn:tauPosterior1}
      \end{equation}
      with
      \begin{equation}
        \textstyle
        g_{\vec{y}}(\tau) \;=\; 
                    \exp\Bigl(-\frac{1}{2} \Bigl[
                      \frac{(y_1-\tilde{\mu}(\tau))^2}{\sigma_1^2+\tau^2}
                      + \frac{(y_2-\tilde{\mu}(\tau))^2}{\sigma_2^2+\tau^2}\Bigr] \Bigr)
        \;=\; \exp\Bigl(-\frac{1}{2}\,
                                     \frac{(y_2-y_1)^2}{\sigma_1^2+\sigma_2^2+2\tau^2}
                              \Bigr) \label{eqn:tauPosterior2}
      \end{equation}
      (see, e.g., Eqn.~(11) in \citet{Roever2020}), where $p(\tau)$~is
      the heterogeneity's prior density, and
      $f_{\vec{\sigma}}(\tau)$~is a lengthier term involving~$\tau$
      and~$\vec{\sigma}$.  From~(\ref{eqn:tauPosterior2}) one can see
      that the heterogeneity's posterior depends on the data ($y_1$,
      $y_2$) only via the absolute difference~$|y_2-y_1|$, which in a
      sense constitutes the ``empirical'' or ``observed'' amount of
      heterogeneity, through the exponential term~$g_{\vec{y}}(\tau)$.
      
      A closer look at~$g_{\vec{y}}(\tau)$ shows that it always
      remains between zero and one ($0 < g_{\vec{y}}(\tau) \leq 1$).
      For $y_2=y_1$, it is constant at $g_{\vec{y}}(\tau)=1$. For a
      given difference $|y_2-y_1|>0$, it takes its minimum at $\tau=0$
      and then increases monotonically with~$\tau$. For any given
      $\tau$ it decreases monotonically in~$|y_2-y_1|$. One might
      think of~$g_{\vec{y}}(\tau)$ as ``ruling out'' smaller
      $\tau$~values in the heterogeneity posterior and pushing the
      posterior mode towards higher $\tau$~values as
      $|y_2-y_1|$~increases.

      The functional form of the posterior (\ref{eqn:tauPosterior2})
      implies that for increasing $|y_2-y_1|$ the resulting marginal
      heterogeneity posterior becomes \emph{stochastically larger}
      \citep{ShakedShanthikumar}; see also the appendix for a
      derivation.
      When varying the prior distribution~$p(\tau)$
      in~(\ref{eqn:tauPosterior1}), we may to some extent also predict
      the effect on the heterogeneity posterior: in particular,
      choosing a stochastically larger heterogeneity prior will imply
      a stochastically larger posterior as well (see also the
      appendix).

    \begin{table}[h]
      \caption{\label{tab:VaryingTau}Effects on several expressions when varying the heterogeneity~$\tau$ between its extremes.}  
      \centering
      \begin{tabular}{llll}
        \toprule
         &&\multicolumn{2}{c}{heterogeneity extreme} \\
        \cmidrule(lr){3-4}
        expression & & $\quad\tau=0$ & $\quad\tau \rightarrow \infty$ \\
        \midrule
        inverse-variance (IV) weight $w_j(\tau)$ & \eqref{eqn:overallMean01} & FE weight $\Bigl(\frac{\sigma_j^{-2}}{\sum_i\sigma_i^{-2}}\Bigr)$ & average weight $\bigl(\frac{1}{k}\bigr)$ \\
        weight $b_j(\tau)$ & \eqref{eqn:shrinkMean01} & $0$ & $1$\\
        shrinkage weight $c_{jj}(\tau)$ & \eqref{eqn:shrinkMean02} & FE weight & $1$ \\[2ex]
        conditional overall mean $\tilde{\mu}(\tau)$ & \eqref{eqn:overallMean01} & FE estimate & unweighted average \\
        conditional shrinkage estimate $\tilde{\theta}_j(\tau)$ & \eqref{eqn:shrinkMean01} & FE estimate & original estimate $y_j$\\
        \bottomrule
      \end{tabular}
    \end{table}
%
%

%%%%%%%%%%%%%%%%%%%%%%%%%%%%%%%%%%%%%%%%%%%%%%%%%%%%%%%%%%%%%%%%%%%%%%%%%%%%%%%%
  \section{Bounds for the study weights}\label{sec:bounds}
    \subsection{Lower bounds}
      The above conditions imply that we can derive bounds for the
      shrinkage weights.  As mentioned previously, concerns are
      sometimes raised that the target estimates may be overwhelmed by
      the source data, i.e., that certain weights may become \emph{too
        small} \citep{WeberHemmingsKoch2018}. In the following, we will
      describe the conditions under which we can derive \emph{lower
        bounds} on weights, i.e., where we can make sure that weights
      remain above a certain minimum.
      Important consequences for the weights, valid quite generally or
      for certain heterogeneity priors~$p(\tau)$, are derived
      below. Note that while we assume the standard errors~$\sigma_i$
      to be given (a common assumption to be made in meta-analysis or
      study design considerations), the data (estimates~$y_i$) or the
      prior ($p(\tau)$) may be varied.

    \subsection{A study's minimum contribution to its own shrinkage estimate: the ``FE~weight''}
      The (conditional) shrinkage weight~$c_{jj}(0)$, i.e. the $j$th
      study's contribution to its own shrinkage estimate evaluated
      at~$\tau=0$, constitutes a lower bound for the posterior mean
      weights.  Any heterogeneity prior~$p(\tau)$ may attach prior
      probability to $\tau$~values larger than zero, for which the
      weights are only increasing. These ``FE~weights'' may simply be
      computed as the common study weights in a fixed-effect
      meta-analysis. This property holds independent of the actual
      data~($y_i$) or the heterogeneity prior~($p(\tau)$).

    \subsection{Minimum posterior mean shrinkage weight: the ``coincidence weight''}
      For any prior distribution~$p(\tau)$, the \emph{coincidence}
      case of $y_1=y_2$ is the data realisation yielding the lowest
      possible posterior mean shrinkage weight. Any data with
      $|y_2-y_1|>0$ will imply a stochastically larger heterogeneity
      posterior that will (due to monotonicity of
      weights~$c_{jj}(\tau)$ as a function of~$\tau$) lead to larger
      posterior mean shrinkage weights. The coincidence weights may
      simply be computed by performing the meta-analysis with the data
      ($y_1$ and $y_2$) substituted by two identical numbers. This
      property holds independent of the data~($y_i$) and for any given
      heterogeneity prior~($p(\tau)$).

    \subsection{Stochastically ordered priors and their posterior mean weights}
      Considering stochastically ordered families of heterogeneity
      priors allows to vary the posterior mean shrinkage weight.  For
      properly chosen stochastically smaller priors, the posterior
      mean may approach the FE~weight, while for stochastically larger
      priors the posterior mean weight may approach 100\%.  An
      obvious, simple way to yield a stochastically ordered family of
      prior distributions for the heterogeneity is by using (or
      introducing) a scale parameter \citep[Sec.~VII.6.2]{MGB}.  An
      example would be given by varying the scale parameter within the
      half-normal family of prior distributions
      \citep{RoeverEtAl2020}.  This property holds for given
      data~($y_i$) and a stochastically ordered family of
      heterogeneity priors~($p(\tau)$).

%%%%%%%%%%%%%%%%%%%%%%%%%%%%%%%%%%%%%%%%%%%%%%%%%%%%%%%%%%%%%%%%%%%%%%%%%%%%%%%%
  \section{Numerical illustration}\label{sec:illustration}
    In order to demonstrate the shrinkage weights' properties, we
    consider an illustrative case motivated by a scenario involving a
    log-odds-ratio (log-OR) endpoint, analogous to the simulation
    scenario discussed by \citet{RoeverFriede2020}. For a study of
    size~$n_i$ featuring two treatment arms and a binary endpoint, the
    results may be summarized in a $2\!\times\!2$ contingency table.
    Assuming an even distribution of events and non-events across
    table cells implies a log-OR estimate with a standard error of
    approximately~$\frac{4}{\sqrt{n_i}}$ \citep{Roever2020}.
    Considering a combination of a ``small'' and a ``large'' study
    with sizes $n_1=25$ and $n_2=400$ then leads to standard errors of
    $\sigma_1=0.8$ and $\sigma_2=0.2$, respectively. We will then
    derive the smallest RCT's shrinkage estimate (for the
    study-specific effect~$\theta_1$) that is of course primarily
    informed by~$y_1$, but supported by the external data~$y_2$.  The
    present case of $\sigma_1 \gg \sigma_2$ is the kind of setting in
    which we expect to see larger gains from shrinkage estimation, and
    this is exactly where using historical data is the most attractive
    in practice.

    For the analysis, we choose a half-normal heterogeneity prior with
    scale~$0.5$ (HN(0.5)), which constitutes a conservative choice for
    the present scenario
    \citep{FriedeRoeverWandelNeuenschwander2017a,RoeverEtAl2020}. For
    illustration purposes, we also utilize a (stochastically larger)
    HN(1.0) prior.  We then fix the target~$y_1$ (arbitrarily) at zero
    and vary the source~$y_2$ in order to investigate the effect on
    the resulting shrinkage estimates and weights.

    \begin{figure*}[t!]
      \begin{center} % cilengths06.R
        \includegraphics[width=0.31\linewidth]{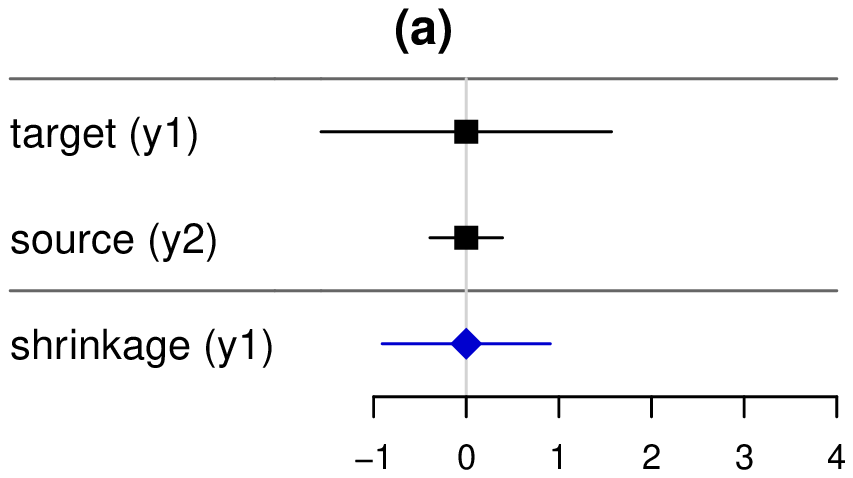}%
        \includegraphics[width=0.31\linewidth]{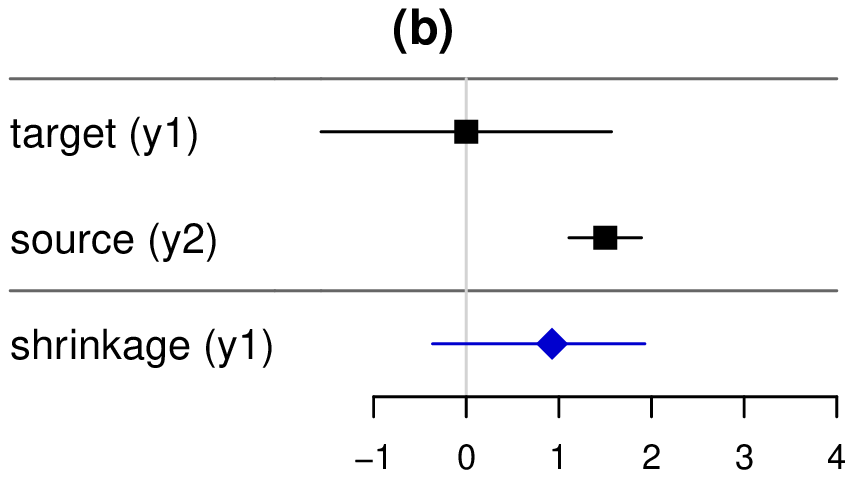}%
        \includegraphics[width=0.31\linewidth]{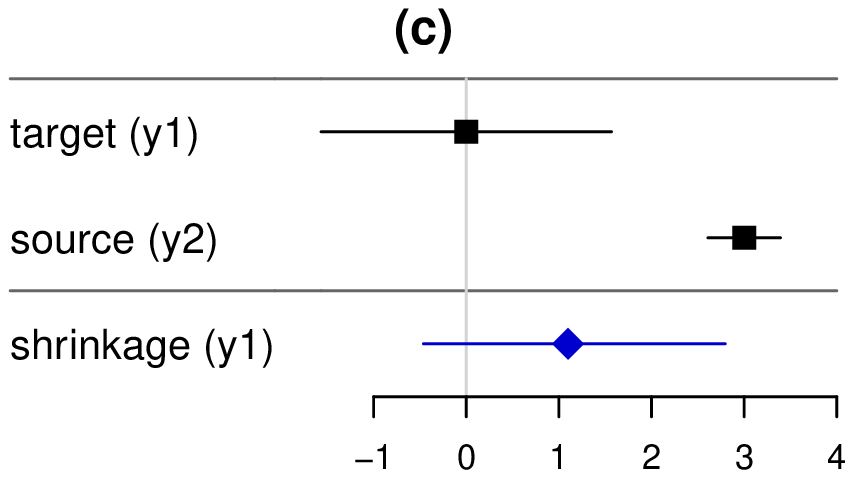} 
        \includegraphics[width=0.95\linewidth]{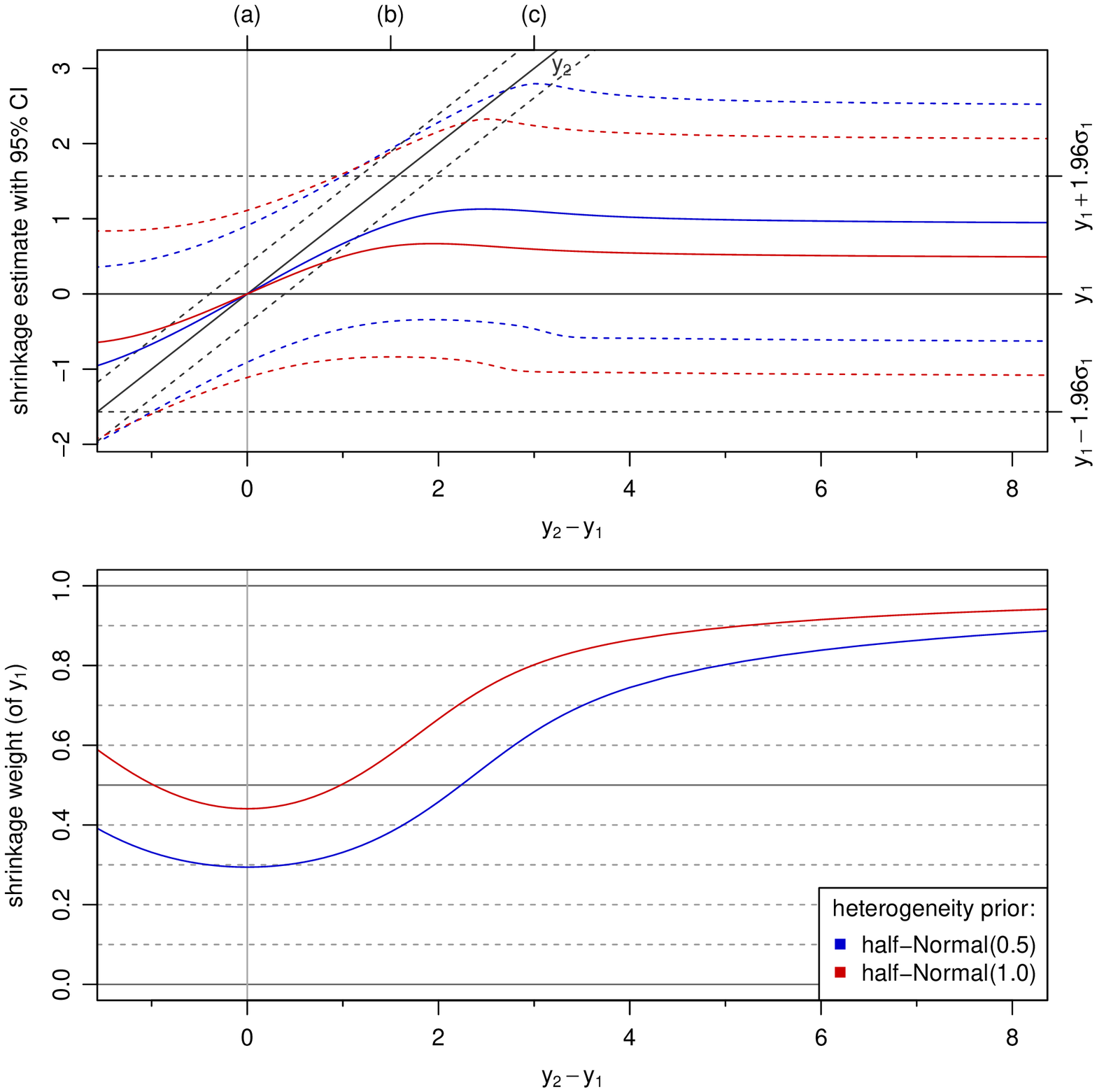} 
        \caption{\label{fig:comparison1} Effect of varying the
          difference between quoted estimates ($y_2-y_1$) on the first
          shrinkage estimate (for~$\theta_1$).  The top row shows
          three data examples of (a) coinciding and (b)--(c) increasingly
          diverging estimates, along with the resulting
          shrinkage estimate for the target study.  The second row
          illustrates the estimates across the continuum of increasing
          $y_2$~values relative to the ``plain'' interval ($y_1 \pm
          1.96 \sigma_1$).
          The bottom panel shows the posterior mean shrinkage weight
          ($\expect[c_{11}(\tau)|\vec{y},\vec{\sigma}]$) for the first
          study, based on two different priors and for varying $y_2-y_1$.
          Note that $y_2-y_1=0$ constitutes the ``coincidence case''.}
      \end{center}
    \end{figure*}

    Fig.~\ref{fig:comparison1} illustrates estimates' and weights'
    dependence on the difference between estimates ($y_1$ and $y_2$).
    The top row of forest plots shows three example cases of
    (a)~coinciding target and source estimates, (b)~some moderate and
    (c)~larger discrepancy between the two; the resulting shrinkage
    estimate for the target is shown in blue.  The second row shows
    the posterior means of~$\theta_1$ (solid lines) and the
    corresponding 95\% CIs (dashed lines) across the continuum of
    source data values. At the top of the plot the three cases
    (a)--(c) are marked, and the blue lines correspond to the
    estimates also shown above. The red lines show analogous
    estimates, but corresponding to a (stochastically larger)
    HN(1.0)~prior.  Note that ``large'' $|y_2-y_1|$~values (here
    e.g. $|y_2-y_1| > 1.96 (\sigma_1 + \sigma_2) = 1.96$) would imply
    non-overlapping CIs for source and target studies (as in
    case~(c)), which in reality may mean that estimates would not
    actually be pooled at all. The practically most relevant bit of
    the plot is hence in the neighbourhood of zero.

    Finally, the bottom plot shows the posterior expected weights to
    illustrate the first (target) study's contribution to its own
    shrinkage estimate.  The minimum (for both heterogeneity priors)
    is attained in the ``coincidence case''~(a) of $y_2-y_1=0$; e.g.,
    for the HN(0.5)~prior the coincidence weight is
    at~$29\%$. Increasing the observed effect difference $|y_2-y_1|$
    (i.e., the ``observed heterogeneity'') then yields increasing
    weights for~$y_1$, implying less borrowing from the source.  In
    cases (b) and (c), the shrinkage weight amounts to $38\%$ and
    $63\%$, respectively.  Also, the choice of a stochastically larger
    prior, here realized by a larger scale parameter in the same
    familiy of distributions, leads to larger weights for~$y_1$, for
    any $|y_2-y_1|$, including the minimum at $|y_2-y_1|=0$.  The
    first study's absolute minimum shrinkage weight, the
    ``FE~weight'', in this case is at
    $c_{11}(0)=\frac{\sigma_1^2}{\sigma_1^2+\sigma_2^2}=\frac{1}{17}=5.9\%$.

    Note that while ``$y_1=y_2$'' constitutes a ``worst case'' in a
    certain sense (leading to the lowest shrinkage weight), it also
    still is the most desirable case, in the sense that this is when
    the data are in agreement and one would expect to learn the most
    from the source study.

%%%%%%%%%%%%%%%%%%%%%%%%%%%%%%%%%%%%%%%%%%%%%%%%%%%%%%%%%%%%%%%%%%%%%%%%%%%%%%%%
  \section{Applications}\label{sec:examples}
  \subsection{Creutzfeldt-Jakob example}
    \begin{table}[b]
      \caption{\label{tab:VargesData} Data from \citet{VargesEtAl2017} on an observational and a
        randomized study investigating the effect of doxycycline on
        survival in Creutzfeldt-Jakob disease (CJD).}  
      \centering
      \begin{tabular}{clcccc}
        \toprule
         &&\multicolumn{2}{c}{patients}&\multicolumn{2}{c}{log(HR)} \\
        \cmidrule(lr){3-4}
        \cmidrule(lr){5-6}
        $i$ & study & treatment & control & $y_i$ & $\sigma_i$\\
        \midrule
        $1$ & observational &           55 &           33 & $-0.499$ & $0.249$ \\
        $2$ & randomized    & \phantom{0}7 & \phantom{0}5 & $-0.173$ & $0.631$ \\
        \bottomrule
      \end{tabular}
    \end{table}
    A small randomized controlled trial (RCT) was conducted in order
    to investigate the effect of doxycycline on survival in patients
    suffering from Creutzfeldt-Jakob disease (CJD)\@. In this ultra
    rare condition, only 12~patients could be recruited, and so data
    from an observational study were considered as complementing
    evidence \citep{VargesEtAl2017}. Both studies quote estimated
    hazard ratios (HRs), and these estimates along with their standard
    errors are jointly analyzed in a meta-analysis; the data are also
    shown in Table~\ref{tab:VargesData}. With the focus being on the
    evidence from the RCT, a shrinkage estimate for this study is
    derived \citep{RoeverFriede2020}.  Both studies are in agreement,
    suggesting a beneficial treatment effect, while the absolute
    effect magnitude is larger for the observational data.

    Since the larger observational study provides a much more precise
    estimate (smaller standard error), one might fear that the
    randomized evidence will be overwhelmed by the external data in a
    joint analysis.  The FE~weight in this case amounts to
    $c_{22}(0)=\frac{\sigma_2^2}{\sigma_1^2+\sigma_2^2}=13.5\%$; this
    would be the RCT's weight in an FE analysis, and it constitutes a
    lower bound on the RCT's weight for any data realization ($y_1$,
    $y_2$) or any heterogeneity prior ($p(\tau)$).

    For a log-HR, we may then assume a half-normal prior with
    scale~0.5 (HN(0.5)) for the heterogeneity
    \citep{FriedeRoeverWandelNeuenschwander2017a,RoeverFriede2020,RoeverEtAl2020}.
    For this prior, we get a minimum posterior mean weight
    (coincidence weight) for the randomized study of~$38.9\%$, which
    may already be considered reassuringly large, in view of the
    sample sizes involved and compared to the FE~weight.  Any data
    realization ($y_1$, $y_2$) will hence yield an eventual weight
    $\geq 38.9\%$ for the RCT\@.  Also, a larger scale of the
    heterogeneity prior (i.e., a larger expected amount of
    heterogeneity) will increase the minimum weight for~$y_2$; e.g., a
    HN(1.0) prior would yield a minimum expected shrinkage weight
    of~$52.1\%$.  For the actual data (Table~\ref{tab:VargesData}), we
    then get a weight of $39.5\%$, slightly above the minimum, for the
    RCT\@.  Table~\ref{tab:VargesEstimates} shows weights and
    estimates corresponding to the two different heterogeneity
    priors. In both cases, the actual weights are not far from their
    minimum value, and for both analyses there is a sizeable gain in
    precision for the shrinkage estimate when compared to the original
    estimate ($y_2$, $\sigma_2$) alone.

    \begin{table} % cilengths06.R
      \caption{\label{tab:VargesEstimates} Estimates for the CJD
        example. For different heterogeneity priors (HN(0.5) or
        HN(1.0)), the corresponding minimum (coincidence) weight is
        given, as well as the resulting weight for the actual data
        along with the corresponding shrinkage estimates.  The very
        last line shows the estimate based only on~$y_2$
        and~$\sigma_2$ for comparison.} 
      \centering
      \begin{tabular}{lcccc}
        \toprule
                 & \multicolumn{2}{c}{mean weight} & \multicolumn{2}{c}{effect estimate~$\theta_2$}\\
        \cmidrule(lr){2-3}  \cmidrule(lr){4-5}
        $\tau$~prior & minimum & actual           & mean & 95\% CI \\
        \midrule
        HN(0.5) & 38.9\% &  \phantom{1}39.5\% & $-0.370$ & [$-1.157$, $0.477$] \\
        HN(1.0) & 52.1\% &  \phantom{1}53.1\% & $-0.326$ & [$-1.232$, $0.664$] \\
        \bottomrule
                  &        & (100.0\%\phantom{)} & $-0.173$ & \phantom{(}[$-1.410$, $1.064$]) \\
      \end{tabular}
    \end{table}

  \subsection{Metabolic acidosis example}
    A gynaecological RCT investigated whether fetal monitoring using
    cardio\-toco\-graphy (CTG) combined with ECG ST-segment analysis
    (ST) reduced the occurrence of metabolic acidosis, compared to CTG
    alone \citep{WesterhuisEtAl2007}. Here the relative risk (RR) of
    metabolic acidosis comparing the two treatment groups is of
    interest.  When analyzing the data, evidence from an earlier,
    similar RCT \citep{AmerWahlinEtAl2001} may be utilized to support
    parameter estimation.  This example data set was originally
    investigated by \citet{RietbergenEtAl2011}; the corresponding data
    are shown in Table~\ref{tab:RietbergenData}.

    \begin{table}[b]
      \caption{\label{tab:RietbergenData} Data from
        \citet{RietbergenEtAl2011} on two gynaecological RCTs
        investigating whether fetal monitoring using
        cardio\-toco\-graphy (CTG) combined with ECG ST-segment
        analysis was associated with a reduced risk of metabolic
        acidosis, compared to CTG alone.}  \centering
      \begin{tabular}{clcccccc}
        \toprule
              && \multicolumn{2}{c}{treatment}
               & \multicolumn{2}{c}{control}
               & \multicolumn{2}{c}{log(RR)} \\
        \cmidrule(lr){3-4}
        \cmidrule(lr){5-6}
        \cmidrule(lr){7-8}
        $i$ & study & events & total & events & total & $y_i$ & $\sigma_i$\\
        \midrule
        $1$ & Amer-W{\aa}hlin (2001) & 15 & 2159 & 31 & 2079 & $-0.764$ & $0.313$ \\
        $2$ & Westerhuis (2007)& 20 & 2827 & 30 & 2840 & $-0.401$ & $0.287$ \\
        \bottomrule
      \end{tabular}
    \end{table}

    Primary interest focuses on the more recent target study by
    \citet{WesterhuisEtAl2007} and on a shrinkage estimate of its
    study-specific effect~$\theta_2$. The two trials are of roughly
    comparable size (5667 vs.\ 4238 participants), and from the
    ``FE~weight'' of
    $c_{22}(0)=\frac{\sigma_2^2}{\sigma_1^2+\sigma_2^2}=54.3\%$ one
    can already see that the second study will definitely contribute
    the majority of weight when estimating its own effect~$\theta_2$.

    For a log-RR, we may again use a half-normal prior with scale~0.5
    for the heterogeneity
    \citep{FriedeRoeverWandelNeuenschwander2017a,RoeverEtAl2020}; this yields a
    minimum (coincidence) mean shrinkage weight of $72.5\%$.  A larger
    heterogeneity prior scale again leads to an increased shrinkage
    weight; e.g., for a HN(1.0) prior, the minimum weight is at
    $78.7\%$.
    \begin{table}   % cilengths06.R
      \caption{\label{tab:RietbergenEstimates} Estimates for the
        metabolic acidosis example. For different heterogeneity priors
        (HN(0.5) or HN(1.0)), the corresponding minimum (coincidence)
        weight is given, as well as the resulting weight for the
        actual data along with the corresponding shrinkage estimates.
        The very last line shows the estimate based only on~$y_2$
        and~$\sigma_2$ for comparison.}  \centering
      \begin{tabular}{lcccc}
        \toprule
                 & \multicolumn{2}{c}{mean weight} & \multicolumn{2}{c}{effect estimate~$\theta_2$}\\
        \cmidrule(lr){2-3}  \cmidrule(lr){4-5}
        $\tau$~prior & minimum & actual           & mean & 95\% CI \\
        \midrule
        HN(0.5) & 72.5\% &  \phantom{1}74.0\% & $-0.495$ & [$-0.986$, $0.005$] \\
        HN(1.0) & 78.7\% &  \phantom{1}80.5\% & $-0.472$ & [$-0.983$, $0.051$] \\
        \bottomrule
                  &        & (100.0\%\phantom{)} & $-0.401$ & \phantom{(}[$-0.964$, $0.163$]) \\
      \end{tabular}
    \end{table}
    Table~\ref{tab:RietbergenEstimates} shows the corresponding
    weights and estimates. Compared to the previous example, the
    precision gain is not quite as large here.

%%%%%%%%%%%%%%%%%%%%%%%%%%%%%%%%%%%%%%%%%%%%%%%%%%%%%%%%%%%%%%%%%%%%%%%%%%%%%%%%
  \section{Conclusions}\label{sec:conclusion}

    Bayesian meta-analysis provides a transparent means for
    extrapolation or \emph{borrowing of strength} from external data
    \citep{RoeverFriede2020}.
    Also within a Bayesian inference framework, study weights for
    overall and shrinkage effect estimates may be derived as posterior
    expected weights, for any number of studies~$k$\@.  The FE~weights
    (conditional on $\tau\!=\!0$) constitute the absolute minimum
    shrinkage weights across all heterogeneity priors and data
    realizations.
    In case of $k\!=\!2$ studies, the heterogeneity posterior depends
    on the data only via the absolute difference in both estimates
    ($|y_2-y_1|$).  A larger difference leads to a stochastically
    larger heterogeneity posterior. When the estimates coincide,
    i.e. $y_2=y_1$, the smallest possible shrinkage weight for a given
    heterogeneity prior (across all possible data realizations) is
    obtained.
    Concerning the choice of heterogeneity prior, a stochastically
    larger prior leads to a stochastically larger posterior, and with
    that to increased (minimum and actual) shrinkage weights.

    The above findings have important implications for the weightings
    that may occur within a meta-analysis.  The shrinkage weight is
    bounded below (irrespective of the prior and data) by the
    FE~weight.  For any particular given prior, the (posterior mean)
    shrinkage weight is also bounded below across possible data
    realisations by the ``coincidence weight''.  Having a bound on the
    weight effectively means bounding the ``leverage'' of the external
    data for the shrinkage estimate. A lower bound of, say, 50\% means
    that the resulting shrinkage estimate will not move more than
    halfway from the effect~$y_i$ towards the external data in case of
    near-concordant evidence. For greater discrepancies, the target
    study's weight will even be larger, or conversely, the influence
    of the source data will be smaller.

    The FDA Guidance on ``Leveraging existing clinical data for
    extrapolation to pediatric uses of medical
    devices''\citep{FDA2016} for example elaborates on issues commonly
    encountered in extrapolation endeavours. One concern raised here
    is the exchangeability assumption~(\ref{eqn:NNHM3}) commonly made
    in hierarchical models. In the common case of only $k\!=\!2$
    studies, however, the same model (as far as shrinkage estimation
    is concerned) may alternatively be motivated via the
    \emph{reference model} \citep{RoeverFriede2020}. This is similar
    to the \emph{bias allowance model} framework \citep{WeltonEtAl},
    where the target study is estimating the parameter of interest
    ``directly'', while the source is associated with a potential bias
    term of unknown direction and magnitude. Moreover, the advantages
    of using (informative) priors on the heterogeneity parameter are
    acknowledged in the guidance document, in particular as this
    facilitates dynamical borrowing based on the empirically observed
    compatibility of source and target data.

    We would like to encourage consideration of minimum weights as a
    \emph{diagnostic tool} of the evidence constitution and of
    implications of prior settings for a given or anticipated data
    scenario.  The study of weights should, however, not be used for
    guiding the selection of the heterogeneity prior. The choice of
    prior should primarily be driven by considerations of prior
    information on between-study variability. Different amounts on
    heterogeneity might however be anticipated in different contexts
    --- e.g., in the two examples discussed above, greater
    heterogeneity may be plausible between observational and
    randomized data than for the case of two RCTs
    \citep{RoeverEtAl2020}.

    A closely related approach to investigating the contributions of
    data sources to a joint analysis works via the consideration of
    \emph{effective sample sizes (ESS)}
    \citep{NeuenschwanderEtAl2020}. Target and source data may be
    assessed based on their ``share'' of the total sample size, and
    also the effect of heterogeneity on the resulting MAP prior may be
    quantified via \emph{prior maximum sample sizes}
    \citep{NeuenschwanderEtAl2010}. Robustification of a MAP prior may
    be achieved by implementing a mixture of vague and informative
    prior components \citep{SchmidliEtAl2014}. In case the source
    data's weight is considered \emph{too large}, a simple remedy
    might be to artificially inflate its associated standard error,
    similarly to the idea behind a \emph{power prior} approach
    \citep{IbrahimChen2000}. Alternatively, the target sample size
    might also be increased, in case that is an option. For an
    overviev of approaches to downweighting of external data, see also
    \citet{VieleEtAl2014}.

    The considerations which provide some insights into the inner
    workings of shrinkage estimation facilitate diagnostics even
    before considering actual data.  A requirement is that the
    standard errors~($\sigma_i$) need to be known beforehand. Quite
    often these may be approximated based on a \emph{unit information
      standard deviation (UISD)} and the sample size
    \citep{KassWasserman1995,RoeverEtAl2020}. The need to make
    assumptions about anticipated standard errors is a common issue in
    similar design-of-experiment contexts \citep{Cohen}.  Fears of
    external evidence ``overruling'' the target data
    \citep{WeberHemmingsKoch2018} may be unwarranted, or may be
    checked before carrying out the target study, as the NNHM behaves
    predictably and reasonably within a Bayesian analysis.  Potential
    problems arise or are amplified when using frequentist methods:
    the concerningly common occurrence of zero heterogeneity estimates
    means that analyses may fall back to an FE~approach, which here is
    the least cautious or least conservative analysis.  For the case
    of few studies, the probability of obtaining a zero heterogeneity
    estimate is alarmingly high --- approaching 50\% even for moderate
    amounts of heterogeneity
    \citep{FriedeRoeverWandelNeuenschwander2017a}, which may actually
    render frequentist heterogeneity estimation for small~$k$ a
    somewhat questionable exercise.  Within a Bayesian framework,
    marginalisation over the plausible range of heterogeneity values
    will lead to a more conservative behaviour.  In summary, with the
    target study's contribution to the resulting Bayesian shrinkage
    estimate being bounded below, concerns of evidence being easily
    overwhelmed by external source data can be addressed
    a\mbox{-}priori, and may be shown to be largely unwarranted.

\section*{Acknowledgment}
  Support from the \emph{Deutsche Forschungsgemeinschaft (DFG)} is
  gratefully acknowledged (grant number \mbox{FR~3070/3-1}).

\section*{Conflicts of interest}
  The authors have declared no conflict of interest.

\section*{ORCID}
  Christian R\"{o}ver: \href{http://orcid.org/0000-0002-6911-698X}{0000-0002-6911-698X}\\
  \noindent
  Tim Friede: \href{http://orcid.org/0000-0001-5347-7441}{0000-0001-5347-7441}

%%%%%%%%%%%%%%%%%%%%%%%%%%%%%%%%%%%%%%%%%%%%%%%%%%%%%%%%%%%%%%%%%%%%%%%%%%%%%%%%
\clearpage
\appendix
  \section{Appendix}
  \subsection{Stochastic ordering of heterogeneity posteriors}\label{sec:dataOrderAppendix}
    Consider two parameter sets $\vec{y}_a$ and $\vec{y}_b$ for which
    $0 \leq |y_{a;2}-y_{a;1}| < |y_{b;2}-y_{b;1}|$. Then the ratio of
    the heterogeneity's marginal posterior densities is given by
    (cf.~(\ref{eqn:tauPosterior1}))
    \begin{equation}
      \frac{p(\tau|\vec{y}_b,\vec{\sigma})}{p(\tau|\vec{y}_a,\vec{\sigma})}
      \;=\; \frac{c_{\vec{y}_b} \, p(\tau) \, g_{\vec{y}_b}(\tau)}{c_{\vec{y}_a} \, p(\tau) \, g_{\vec{y}_a}(\tau)}
      \;=\; \frac{c_{\vec{y}_b}}{c_{\vec{y}_a}} \;
            \frac{g_{\vec{y}_b}(\tau)}{g_{\vec{y}_a}(\tau)}
      \;\propto\; \frac{g_{\vec{y}_b}(\tau)}{g_{\vec{y}_a}(\tau)}
      \mbox{,}
    \end{equation}
    where $c_{\vec{y}_a}$ and $c_{\vec{y}_b}$ are the densities'
    normalizing constants, and the where the latter ratio of
    ``$g_{\vec{y}}(\tau)$'' terms is monotonically increasing
    in~$\tau$. With that, condition~(C) in \citet{Lehmann1955} is
    fulfilled, and the posterior corresponding to~$\vec{y}_b$ is
    \emph{stochastically larger} than the one associated
    with~$\vec{y}_a$.

  \subsection{Stochastic ordering of posteriors for different priors}\label{sec:priorOrderAppendix}
    Consider two heterogeneity priors with densities~$p_1(\tau)$
    and~$p_2(\tau)$ where $p_2$~is stochastically larger than~$p_1$.
    A posterior distribution constitutes a special case of a
    ``weighted distribution'' \citep{Meczarski2015}. For the posterior
    distributions corresponding to~$p_1$ and~$p_2$ follows that these
    will inherit the same stochastic ordering
    \citep{BartoszewiczSkolimowska2006}.

  \subsection{\textsf{R}~code for CJD example}
{ \footnotesize 
%\verbatiminput{examplecode-CJD.R} 
\begin{verbatim}
# specify data:
cjd <- cbind.data.frame("study"   =c("observational", "randomized"),
                        "logHR"   =c(-0.499, -0.173),
                        "logHR.se"=c(0.249, 0.631))

# analyze:
library("bayesmeta")
bm <- bayesmeta(y=cjd$logHR, sigma=cjd$logHR.se, labels=cjd$study,
                tau.prior=function(t){dhalfnormal(t, scale=0.5)})

# show posterior mean shrinkage weights:
bm$weights.theta
# show shrinkage estimates:
bm$theta

# derive FE weights (percentages, using "metafor" library):
weights(rma.uni(yi=cjd$logHR, sei=cjd$logHR.se, slab=cjd$study,
                measure="GEN", method="FE"))
# alternatively, compute directly:
cjd$logHR.se^-2 / sum(cjd$logHR.se^-2)

# determine coincidence (minimum) posterior mean weights:
bayesmeta(y=c(0,0), sigma=cjd$logHR.se, labels=cjd$study,
          tau.prior=function(t){dhalfnormal(t,scale=0.5)})$weights.theta
\end{verbatim}
}

%\clearpage
{
  \bibliographystyle{bimj}
  \bibliography{../../literature/literature}
}
\end{document}